\documentclass[%
 aps,
 prl,
 amsmath,amssymb,
 longbibliography,
 reprint,
 twocolumn,
]{revtex4-2}
\usepackage{graphicx}
\usepackage[dvipsnames]{xcolor}
\usepackage{hyperref,url}
\usepackage[capitalise]{cleveref}
\hypersetup{breaklinks,colorlinks=true,linkcolor=NavyBlue,citecolor=NavyBlue,urlcolor=NavyBlue,filecolor=NavyBlue}

\makeatletter
\renewcommand{\section}{\@startsection{section}{3}{\z@}{3.25ex \@plus 1ex}{-1em}{\fontsize{11}{13}\selectfont\bfseries}}
\renewcommand{\paragraph}{\@startsection{subsection}{4}{\z@}{3.25ex \@plus 1ex}{-1em}{\bfseries}}
\newcounter{SIref}
\newcounter{SIcite@cnt}
\newcommand{\citeS}[1]{%
  \textsuperscript{%
  \setcounter{SIcite@cnt}{0}%
  \@for\SIcite@tmp:=#1\do{%
    \ifnum\value{SIcite@cnt}>0{,}\fi
    \stepcounter{SIcite@cnt}%
    S\ref{SIref:\SIcite@tmp}}}%
}

\makeatother

\begin{document}

\title{First-Principles Origins of Charge Transport in Molecular Semiconductors}

\author{Tong Jiang}
\author{Joonho Lee}
\email{joonholee@g.harvard.edu}
\affiliation{Department of Chemistry and Chemical Biology, Harvard University, 12 Oxford St, Cambridge, Massachusetts 02138, USA}

\begin{abstract}
Charge transport governs organic transistors and photovoltaics, yet predicting it from atomic structure remains challenging.
Electron--phonon interactions span disparate frequencies, strengths and spatial ranges, and collectively generate nonperturbative carrier dynamics.
Existing methods regain tractability only by assuming a mechanism or reducing electron--phonon coupling to a few modes.
We introduce a parameter-free framework that instead computes transport from \textit{ab initio} electron--phonon Hamiltonians, propagating carriers across hundreds-of-molecule domains with the full phonon spectrum and letting transport regimes and bottlenecks emerge from nonperturbative Green--Kubo dynamics.
Across five representative crystals, it captures measured mobilities, temperature exponents, and optical-conductivity fingerprints.
Our results overturn the prevailing microscopic mechanism for DNTT, tracing its transient localization to correlated on-site disorder from acoustic phonons rather than independent hopping fluctuations.
The resulting two-axis transport map provides design principles and highlights the underexplored phenacene family, exemplified by the high-mobility picene, as a promising direction.
\end{abstract}

\maketitle

Charge transport in molecular semiconductors underlies optoelectronic technologies including flexible and lightweight transistors, photovoltaics, and light-emitting devices~\cite{zhangTwodimensionalPolyanilineCrystal2025,claesPhononlimitedElectronicTransport2025,xie2024high,phillips2022stranger,luMetallicChargeTransport2026}. In these materials, charge motion occurs through weakly bound molecular domains whose electronic couplings, lattice fluctuations, and structural disorder are strongly intertwined~\cite{franchini2021polarons}.
Predicting charge transport from atomic structure is therefore challenging because thermal lattice fluctuations over broad frequency and momentum ranges couple strongly to electronic motion, placing carriers in a nonperturbative regime that is neither well described by band transport nor by hopping between molecules~\cite{fratini2017map,oberhofer2017charge,li2021general}. In this regime, small chemical or packing changes can strongly reshape mobility, optical response, and carrier coherence~\cite{ma2026physical,biswasExcitonPolaronFormation2024,bronstein2020role,fratini2020charge,gianniniChargeTransportOrganic2022,schweicherChasingKillerPhonon2019}.

Phenomenological models, chemically motivated heuristics, and increasingly realistic atomistic simulations have each provided valuable insight into charge-carrier dynamics
~\cite{nelsonModelingChargeTransport2009,shuai2014charge,shuai2020applying,giannini2019quantum,xi2012first,changIntermediatePolaronicCharge2022,zhengAbInitioRealtime2023,fratiniTransientLocalizationScenario2016}.
However, most existing approaches gain tractability by prescribing a particular transport mechanism (band transport~\cite{changBandlikeChargeTransport2024}, molecular hopping~\cite{shuai2014charge,shuai2020applying,nelsonModelingChargeTransport2009}, or transient localization~\cite{fratiniTransientLocalizationScenario2016,fratini2017map}), reducing electron--phonon interactions to a few effective modes~\cite{li2021general}, or describing lattice dynamics with empirical force fields~\cite{giannini2019quantum}.
These choices enable material-specific calculations but sacrifice aspects of the chemical and vibrational complexity of the solid-state transport problems. Consequently, the microscopic origin of the operative transport regime and its controlling bottlenecks can remain elusive.

\begin{figure*}[htbp]
  \centering
  \includegraphics[width=0.96\textwidth]{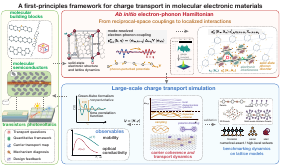}
  \caption{
    \textbf{A first-principles framework for charge transport in molecular electronic materials.}
    Starting from molecular building blocks and crystal structures, periodic electronic-structure and lattice-dynamics calculations provide mode-resolved electron--phonon couplings.
    A gauge-invariant localization procedure transforms these couplings into compact real-space interactions.
    The resulting localized Hamiltonian enables large-scale nonperturbative simulations of real-space carrier coherence and transport dynamics in domains containing up to 648 molecules and more than 160,000 phonon modes.
    The dynamics module is benchmarked against numerically exact or high-level reference calculations on model Hamiltonians before the full framework is applied to real systems with first-principles Hamiltonians.
    The full framework yields quantitative mobility and optical-response predictions, a two-axis transport map, microscopic bottleneck diagnosis, and coherence-guided design feedback for molecular semiconductors relevant to transistors and photovoltaics.
}
  \label{fig:lattice} 
\end{figure*}

What has been missing is a framework that builds solid-state \textit{ab initio} electron--phonon Hamiltonians and lets the operative transport regime and its bottlenecks emerge from nonperturbative dynamics rather than being prescribed in advance.
Here we introduce such a framework (Fig.~\ref{fig:lattice}).
It propagates carriers across domains of hundreds of molecules that retain the full solid-state phonon spectrum, a scale not previously accessible, and reproduces measured mobilities, temperature dependences, and optical-conductivity fingerprints across representative molecular crystals.
%
It resolves why some materials conduct band-like while others localize, sorting them along two physical coordinates: electronic connectivity and low-frequency electron--phonon disorder. The resulting transport map clarifies when established heuristics apply and how molecular structure protects or destroys carrier coherence.
It overturns the accepted bottleneck of the archetypal low-mobility crystal DNTT, revealing a new microscopic origin for its transient localization.
As a design rule, it points to the phenacene semiconductors, with picene as a high-mobility case in point, as a chemically accessible class that has gone largely underexplored.

\section{Results}
\paragraph{Overview of the first-principles framework.}

Our computational strategy is summarized in Fig.~\ref{fig:lattice}, which connects \textit{ab initio} electron--phonon Hamiltonians to domain-scale transport simulations and mechanistic diagnostics.
Starting from molecular structures and crystal packing, periodic electronic-structure and lattice-dynamics calculations yield electronic bands, phonons, and mode-resolved electron--phonon matrix elements~\cite{RevModPhys.89.015003,perturbo}.
We develop a gauge-invariant localization procedure that transforms the reciprocal-space electron--phonon Hamiltonian into a compact real-space representation. The localized Hamiltonian retains the full content of solid-state phonon bands, while making onsite disorder, hopping modulation, and correlations among these phonon-induced fluctuations explicit. This representation connects first-principles inputs to nonperturbative real-space dynamics within the Green--Kubo formalism~\cite{kubo2012statistical} and enables the mechanistic and carrier-coherence studies below.
The dynamical simulation combines a variational polaron treatment of dispersionless high-frequency modes with classical propagation of thermally occupied dispersive low-frequency modes~\cite{fetherolf2020unification,runesonChargeTransportOrganic2024,*runeson2025nuclear} and uses a memory-kernel treatment of the current response~\cite{mori1965,PhysRevLett.114.086601} to efficiently approach thermodynamic and long-time limits.
Details are provided in Methods and Supplementary Materials.

\begin{figure*}
  \centering 
  \includegraphics[width=\textwidth]{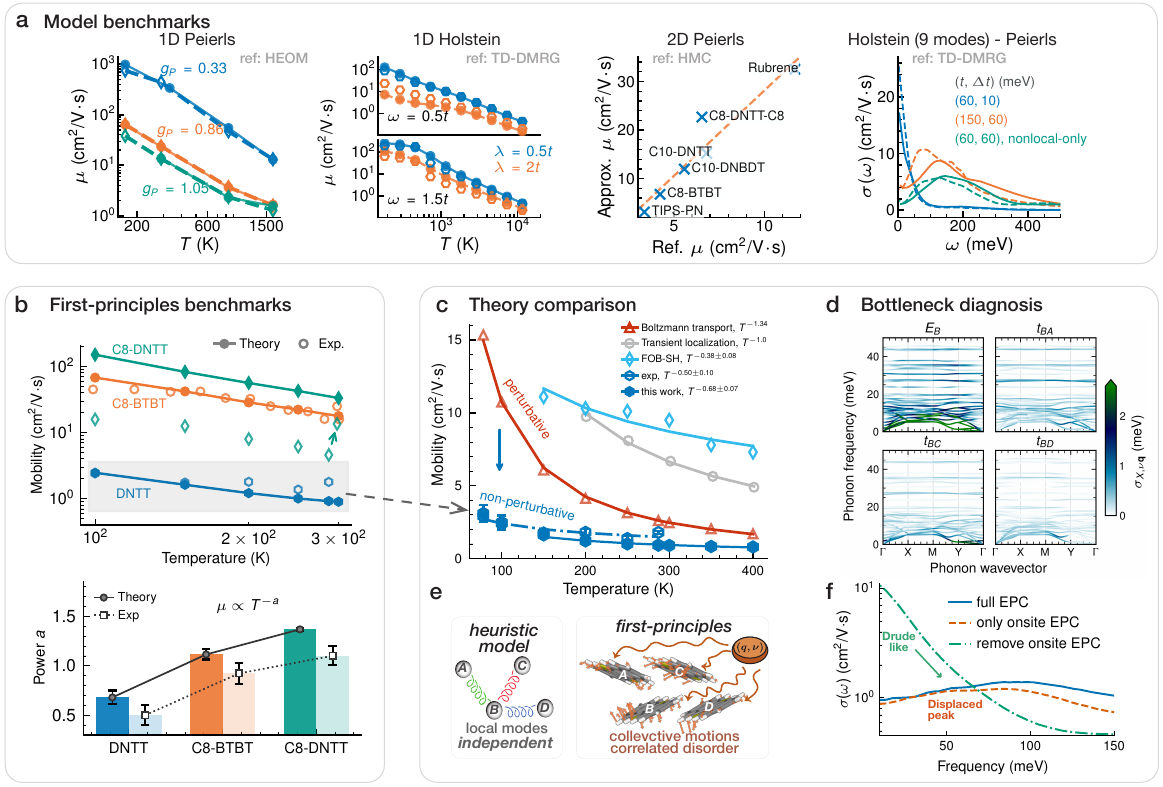}
  \caption{
    \textbf{Benchmarked first-principles dynamics reveal microscopic transport bottlenecks.}
      \textbf{(a)} Dynamics benchmarks on model electron--phonon Hamiltonians, comparing against numerically exact or high-level reference results for one-dimensional Peierls and Holstein, two-dimensional Peierls, and multimode Holstein--Peierls models.
      \textbf{(b)} First-principles mobility benchmarks against experiments.
      Solid and open symbols denote simulations and experimental values, respectively. 
      Field-effect transistor (FET) data are shown for C8-BTBT~\cite{cho2015band,yuanUltrahighMobilityTransparent2014}; DNTT and C8-DNTT mobilities are extracted from optical-pump THz-probe spectroscopy~\cite{gianniniTransientlyDelocalizedStates2023}; the arrow for C8-DNTT marks a recent conductive atomic force microscopy measurement on FET devices~\cite{giceviciusProbingOutOfPlaneCharge2025}. Power-law exponents are obtained from fits to $\mu \propto T^{-a}$.
      \textbf{(c)} DNTT comparison with representative prior approaches. The present nonperturbative calculation is compared with experiment, the perturbative Boltzmann transport equation, transient localization theory, and atomistic fragment-orbital-based surface hopping.
      \textbf{(d)} Mode- and momentum-resolved phonon-induced disorder in DNTT for the onsite energy $E_{\mathrm B}$ and hopping channels $t_{\mathrm{BA}}$, $t_{\mathrm{BC}}$, and $t_{\mathrm{BD}}$, where $A$, $B$, $C$, and $D$ denote the molecular sites shown in \textbf{(e)}.
      \textbf{(e)} Schematic contrast between independent local-mode heuristics and the first-principles solid-state picture, where a single solid-state phonon mode can collectively modulate several electronic channels in a correlated manner.
      \textbf{(f)} Optical conductivity under channel interventions.
      }
  \label{fig:dntt}  
\end{figure*}

\begin{figure*}
  \centering 
  \includegraphics[width=\textwidth]{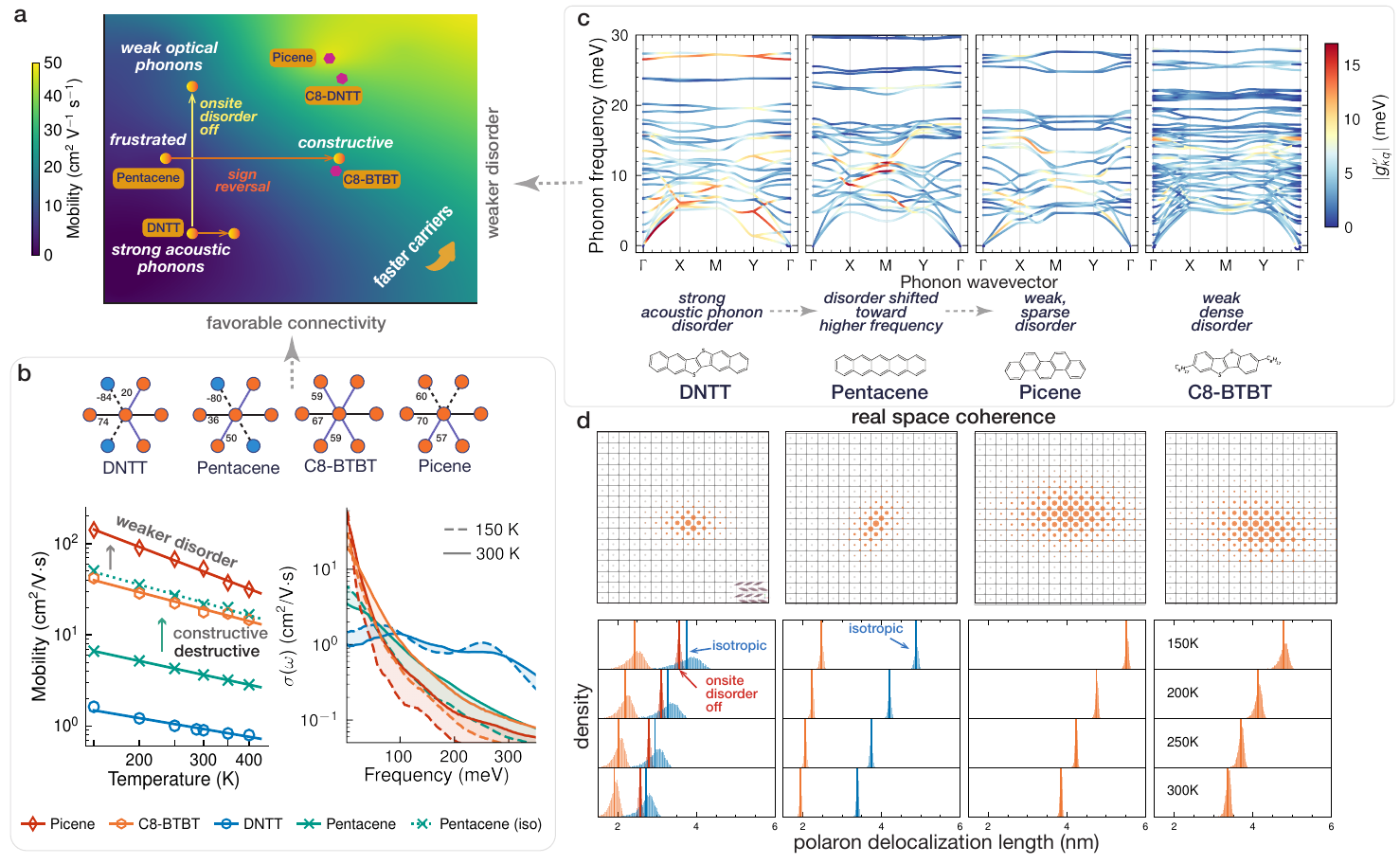}
  \caption{
    \textbf{A first-principles transport map links electron--phonon disorder, electronic connectivity, and carrier coherence.}
    \textbf{(a)} First-principles transport map defined by electron--phonon disorder and electronic connectivity. Moving upward corresponds to weaker disorder, whereas moving rightward corresponds to more favorable connectivity and constructive interference.
    The continuous mobility background provides a visual interpolation between explicitly computed points.
    The axis positions encode relative rankings inferred from the calculated electron--phonon spectra and signed transfer-integral networks.
    \textbf{(b)} Transfer-integral motifs (values in meV) illustrate frustrated and constructive connectivity, while mobility trends and optical conductivity spectra show the progression from transiently localized DNTT transport toward higher mobility and a more Drude-like optical response in weakly disordered, constructively connected materials.
    \textbf{(c)} Phonon dispersions overlaid with mode-resolved electron--phonon coupling strengths.
    \textbf{(d)} Carrier delocalization and real-space coherence across the same material sequence.
    Upper panels show the corresponding thermally averaged carrier coherence at 300 K.
    Lower panels show temperature-dependent delocalization-length distributions for each material's unmodified Hamiltonian (orange) and, where shown, controlled interventions that remove onsite disorder (red) or isotropize the electronic connectivity (blue).
    Vertical lines indicate the mean delocalization length.
  }
  \label{fig:pentacene_DNTT}  
\end{figure*}

\paragraph{Validation of the first-principles framework.}
Our framework is validated at two levels.
First, the approximate Green--Kubo dynamics reproduce reference mobilities and optical conductivities obtained from numerically exact or high-level calculations for one-dimensional Peierls~\cite{jankovic2025charge1} and Holstein, two-dimensional Peierls, and multimode Holstein--Peierls models (Fig.~\ref{fig:dntt}a). 
The two-dimensional Peierls model simulations use parameters of several molecular semiconductors~\cite{PhysRevApplied.22.L031004}, whereas the multimode Holstein--Peierls model uses nine intramolecular modes parameterized for rubrene~\cite{li2021general}.
Second, we apply the full first-principles workflow to representative molecular semiconductors and compare the resulting transport observables with experimental measurements (Fig.~\ref{fig:dntt}b).
The calculations quantitatively capture both the experimental mobilities and the overall temperature dependences.
The mobility agreement is remarkably good for DNTT and C8-BTBT, whereas the calculated intrinsic mobility of C8-DNTT exceeds that inferred from optical-pump THz-probe measurements~\cite{gianniniTransientlyDelocalizedStates2023}. A recent study combining conductive atomic force microscopy with transistor characterization reported substantially higher field-effect mobility~\cite{giceviciusProbingOutOfPlaneCharge2025}, narrowing this gap.

A first-principles calculation isolates the intrinsic single-crystal-domain limit, distinct from the extrinsic factors (contacts, grain boundaries, morphology, and other probe-dependent effects) that introduce uncertainty in measured mobilities~\cite{bittle2016mobility,giceviciusProbingOutOfPlaneCharge2025,hwang2026measuring,takeyama2012organic,*jurchescuEffectImpuritiesMobility2004,*ostroverkhova2005bandlike,*minariTemperatureElectricfieldDependence2006}.
The calculated mobility magnitudes and temperature trends therefore characterize the intrinsic single-crystal transport of each material. They are not fitted to any individual experiment but provide a reference that measurements on increasingly well-controlled samples are expected to approach.
We corroborate this reference through the independent cross-checks below.
Independent photoemission benchmarks show that the calculated bandwidths and transfer integrals agree with angle-resolved photoemission spectroscopy measurements (Supplementary Fig.~\ref{fig:transfer_integrals}).
As shown in later sections, a complementary mechanistic consistency check comes from C8-DNTT and C8-BTBT: despite similar alkylated $\pi$-systems and transfer-integral networks, weaker electron--phonon coupling in C8-DNTT is consistent with its higher intrinsic mobility (Supplementary \cref{fig:dntt_btbt}).
Together, these comparisons support the framework as a controlled reference for intrinsic crystal-domain transport and motivate the mechanistic analysis that follows.

\paragraph*{Collective vibrations as the transport bottleneck.}

DNTT serves as a stringent case for microscopic diagnosis, given the quantitative agreement between the calculated and experimental mobilities.
Its transport combines low mobility, unusually weak temperature dependence, and a displaced-Drude optical response, while representative prior approaches yield substantially different predictions (Fig.~\ref{fig:dntt}c).
The perturbative Boltzmann transport equation (BTE)~\cite{perturbo},
evaluated using the same first-principles Hamiltonian, fails quantitatively, indicating that a weak-scattering description of extended band states is catastrophically wrong.
Transient localization theory~\cite{fratiniTransientLocalizationScenario2016,gianniniTransientlyDelocalizedStates2023}
does not reproduce the observed temperature dependence, whereas fragment-orbital-based surface hopping (FOB-SH) captures the temperature trend through quantum carrier dynamics along atomistic trajectories using empirical force fields and analytically approximated transfer integrals~\cite{gianniniTransientlyDelocalizedStates2023}.
This provides an all-atom-level description of dynamic disorder, while its attribution to collective solid-state phonons and their electronic action remains indirect.
The channel- and momentum-resolved analysis identifies the bottleneck directly (Fig.~\ref{fig:dntt}d). This confirms the long-standing view that low-frequency nonlocal phonons control dynamic disorder~\cite{PhysRevLett.96.086601,fratiniTransientLocalizationScenario2016,fratini2017map,schweicherChasingKillerPhonon2019,changBandlikeChargeTransport2024}, but changes the microscopic assignment.
The leading disorder does not consist simply of independent hopping fluctuations between neighboring molecules; instead, it is correlated solid-state disorder with a dominant onsite component.
A single low-frequency phonon mode can therefore modulate onsite energies and several hopping pathways with correlated signs (Supplementary Fig.~\ref{fig:acoustic}c).
The relevant modes are not uniform $\Gamma$-point translations, but thermally populated finite-$\mathbf q$ acoustic distortions that combine relative molecular displacements and local crystal-field changes.
This collective effect is obscured in previous monomer- and dimer-based descriptions that treat local modes independently~\cite{sanchez2010theoretical,troisiChargeTransportHigh2011,shuai2014charge,fratiniTransientLocalizationScenario2016,giannini2019quantum} (Fig.~\ref{fig:dntt}e).

In addition, the optical response provides a complementary dynamical signature of this bottleneck.
Both the full-coupling and onsite-only calculations produce a displaced-Drude peak and similar lineshapes, whereas removing the onsite channel restores a Drude-like low-frequency response (Fig.~\ref{fig:dntt}f). 
These interventions show that onsite disorder generated by collective acoustic phonons is the dominant factor driving DNTT toward localization.
Because these low-frequency modes remain thermally populated across the experimental temperature range, they generate substantial quasistatic disorder even at low temperature, accounting for the weak temperature dependence of the mobility. 
The analyses below further connect this disorder to reduced carrier delocalization and a pronounced, temperature-insensitive band-edge tail of localized states.

\paragraph*{First-principles comparisons reveal a two-axis transport map.}
Comparison across representative materials reveals a broader organizing principle (Fig.~\ref{fig:pentacene_DNTT}a).
As shown in Fig.~\ref{fig:pentacene_DNTT}b, DNTT and pentacene have negative transfer-integral products around the lattice plaquette, whereas picene and C8-BTBT have positive products.
For hole transport, negative products are associated with frustrated pathways and destructive interference, whereas positive products favor constructive connectivity and higher mobility~\cite{fratini2017map}.
Consistent with this rule, DNTT and pentacene form a lower-mobility group, whereas picene, C8-BTBT, and C8-DNTT share similar transfer networks (Supplementary Fig.~\ref{fig:dntt_btbt}) and form a higher-mobility group.
A direct test confirms the mechanism: reversing the sign of one pentacene transfer integral and its associated electron--phonon couplings substantially increases the mobility (Fig.~\ref{fig:pentacene_DNTT}b).

However, electronic connectivity does not explain the differences within each group.
Pentacene has heavier and more anisotropic valence-band effective masses than DNTT (Supplementary Fig.~\ref{fig:bands}), yet exhibits higher mobility. 
Picene, C8-DNTT, and C8-BTBT have broadly comparable in-plane dispersions (Supplementary Fig.~\ref{fig:bands}), yet exhibit different mobilities. 
Band dispersion alone cannot explain the remaining variation.
The relative disorder coordinate orders the materials by the strength and frequency distribution of electron--phonon couplings, particularly within the thermally populated low-frequency manifold (Fig.~\ref{fig:pentacene_DNTT}c).
DNTT couples strongly to slow, thermally populated acoustic modes that generate a quasistatic disorder landscape, whereas pentacene has weaker coupling shifted toward higher frequencies (Supplementary Fig.~\ref{fig:dntt_pentacene_disorder}). Picene shows weak, sparse low-frequency coupling, whereas C8-BTBT retains weak but denser low-frequency coupling.
The same trend distinguishes the two alkylated systems: C8-DNTT exhibits weaker electron--phonon coupling than C8-BTBT, consistent with its higher mobility (Fig.~\ref{fig:pentacene_DNTT}b and Supplementary Fig.~\ref{fig:dntt_btbt}).
The transport differences reflect the combined action of the two coordinates: mobility rises across the series as connectivity becomes less frustrated and low-frequency disorder weakens, spanning the displaced-Drude response of DNTT to the Drude-like response of the high-mobility materials.

The real-space diagnostics provide a complementary test of this interpretation (Fig.~\ref{fig:pentacene_DNTT}d).
The upper panels show the equilibrium intermolecular carrier coherence, $\langle a_{\mathrm{center}}^\dagger a_j\rangle_{\mathrm{eq}}$, at 300~K. This spatial hierarchy mirrors that inferred from mobility and optical conductivity: coherence is spatially compact in DNTT, broader in pentacene, and more extended in picene and C8-BTBT.
The lower panels further show distributions of the polaron delocalization length from 150 to 300~K. DNTT shows the shortest delocalization lengths but the broadest distributions, reflecting strong configuration-to-configuration variation in the disordered landscape.
Pentacene shows intermediate delocalization, whereas picene and C8-BTBT show distributions shifted toward larger lengths.
Sign-reversal interventions in pentacene and DNTT shift the distributions toward larger values, but the low-delocalization tail persists in DNTT.
Removing onsite disorder in DNTT both increases the delocalization length and substantially suppresses this tail.
The same quasistatic disorder leaves a temporal signature: a time-domain view shows long-lived dynamical memory and prolonged subdiffusion in DNTT, whereas the high-mobility materials exhibit extended superdiffusive windows (Supplementary Fig.~\ref{fig:tcf}b,c).

\begin{figure*}[t]
  \centering
  \includegraphics[width=0.8\textwidth]{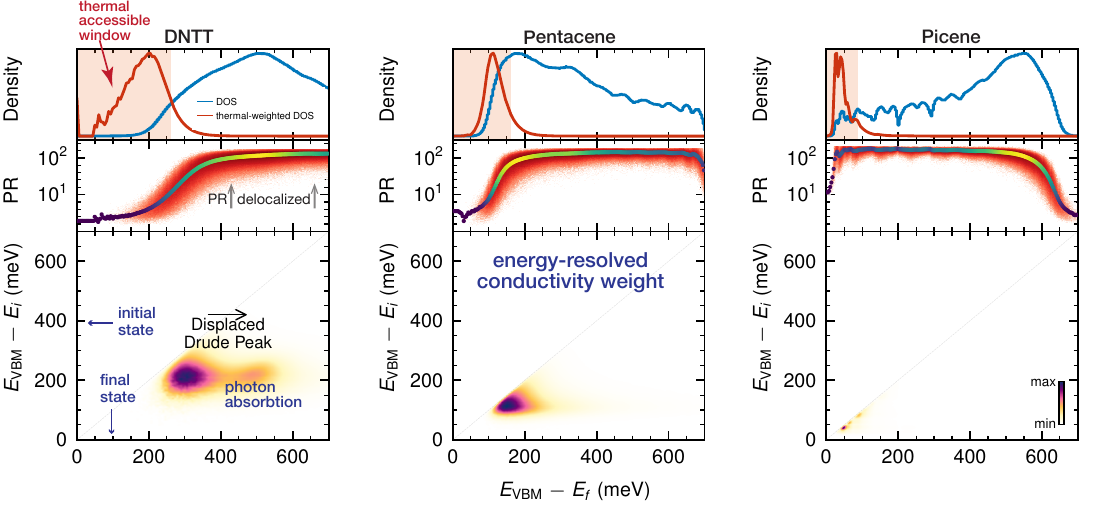}
  \caption{
    \textbf{Energy-resolved localization and optical-conductivity weight reveal how thermal disorder reshapes transport fingerprints.}
    The density of states and its thermally weighted counterpart distinguish available states from those sampled at 300~K.
    The participation ratio (PR) measures spatial delocalization, with larger values indicating more extended states.
    The energy-resolved optical-conductivity weight decomposes the optical response into contributions from transitions between thermally occupied initial states and accessible final states.
}
  \label{fig:polaron}
\end{figure*}

These results connect microscopic Hamiltonian features to mobility, optical response, and domain-scale carrier coherence, placing each material on a two-coordinate map that ranges from the transient-localization regime of frustrated, strongly disordered systems to the Drude-like corner reached only when favorable connectivity and weak dynamical disorder coincide.
Picene occupies this favorable corner and reaches the highest intrinsic mobility among the materials compared here. Its placement is consistent with a recent angle-resolved photoemission measurement linking weak dynamical disorder to the armchair character of its frontier orbitals~\cite{neef2024frontier}.
Because this armchair frontier-orbital topology is a defining structural feature of the phenacene series, the two conditions that place picene in the favorable corner of our map, constructive connectivity together with weak low-frequency disorder, are expected to recur across the family, identifying phenacenes as a structurally motivated transport class rather than a single favorable compound.
Alkali-metal-intercalated picene was also reported as the first hydrocarbon superconductor~\cite{mitsuhashi2010superconductivity}, demonstrating the capability to host delocalized carriers.
Together with recent reports of environmental stability and molecular and interfacial tunability~\cite{jiang2022high}, these observations motivate systematic exploration of charge transport across the underexplored phenacene family.

\paragraph*{Energy-resolved conductivity links localization to transport fingerprints.}

The real-space coherence hierarchy in Fig.~\ref{fig:pentacene_DNTT}d has an energy-space counterpart in the thermally sampled electronic states (Fig.~\ref{fig:polaron}).
Within the thermal window, DNTT develops a pronounced band-edge tail of low-participation-ratio (PR) states, the signature of strong localization.
This tail weakens from pentacene to picene, whose thermally sampled states stay spatially extended, consistent with their placement on the transport map.
The energy-resolved conductivity weight (Eq.~\eqref{eq:Xi}) then resolves the optical conductivity by the energies of the initial and final states, exposing how these localized states shape the response.
In DNTT, the dominant weight connects thermally occupied localized tail states to more delocalized states at higher energy, producing the finite-frequency response underlying the displaced-Drude peak; the characteristic transition energy falls in pentacene and collapses in picene, where the response is dominated by low-energy transitions among delocalized band-edge states.

\paragraph*{Discussion.}

Transfer integrals, lattice fluctuations, and dynamic disorder have long been recognized as the ingredients of charge transport in molecular semiconductors, but in real materials they are entangled across broad ranges of energy and timescale.
Different reduced Hamiltonians and dynamical approximations can therefore yield different mobilities, temperature dependences, and even transport mechanisms for the same material.
It has consequently remained unclear which microscopic features govern a given material, when established heuristics apply, and why structurally similar crystals can exhibit sharply different transport fingerprints.

Our framework resolves this by computing charge transport directly from \textit{ab initio} electron--phonon Hamiltonians at domain scale, without prescribing a transport regime or reducing the problem to a few effective modes.
Calculations of this scope, spanning domains of hundreds of molecules that retain the full solid-state electron--phonon coupling, have not previously been feasible, and they let the operative mechanism emerge from the dynamics rather than being assumed.
The result is a single, self-consistent picture for each material, in which mobility, temperature dependence, optical conductivity, dynamical memory (Supplementary Fig.~\ref{fig:tcf}b), and real-space delocalization are explained together.
It also overturns the accepted microscopic assignment for the archetypal case, DNTT: the dominant disorder is not independent hopping fluctuations between molecules, but correlated, largely onsite disorder driven by collective low-frequency acoustic phonons. This distinction is invisible to local-mode descriptions.

Because the Hamiltonian is derived directly from the solid-state structure and retains both real-space locality and momentum-space information, the framework pinpoints the microscopic bottleneck of a given material while providing a common foundation for more specialized dynamical solvers~\cite{lee2015coherent,song2015new,lian2019non,wang2020surface,ren2022time,wu2025nonadiabatic,baumgarten2026scalable}.
In doing so, it turns charge transport in molecular materials from an empirical search into a microscopic, predictive one. Carrier coherence has been proposed as a design principle for charge and energy transport in molecular systems and photovoltaics~\cite{Scholesusing,bredas2017photovoltaic}, and our results show concretely how material-specific interactions preserve or suppress that coherence at the molecular-domain level.

The principal cost is the underlying phonon and electron--phonon calculation; machine-learning surrogates~\cite{zhong2024accelerating,gu2024deep} and semiempirical methods~\cite{bavcic2020analytical} could remove this bottleneck and turn the transport map into a high-throughput screen for new molecular semiconductors. More broadly, the same route extends to other soft semiconductors~\cite{biswasExcitonPolaronFormation2024,fu2025fundamentals,tjheNonequilibriumTransportPolymer2024,aydin2024quantum} and to disorder and interfaces~\cite{bennecke2025hybrid}, doping~\cite{liangNtypeChargeTransport2021}, and anharmonic dynamics~\cite{asher2020anharmonic,banksUntanglingFundamentalElectronic2023}, pointing toward predictive, material-specific transport theory across molecular electronics.

\paragraph*{Acknowledgments.}
The authors thank Lachlan Lindoy, Todd Martinez, David Reichman, Tim Berkelbach, Jiajun Ren, Weitang Li, and the Lee group members for helpful discussions.
The authors acknowledge Zeyu Zhou for his early-stage exploratory work.
This work was funded by Harvard University’s
startup funds.
T.J. acknowledges the Gordon and Betty Moore Foundation Fellowship.
This work used computational resources from FASRC supported by the FAS Division of Science Research Computing Group at Harvard, and the Delta system at the National Center for Supercomputing Applications through allocation CHE250116 from the Advanced Cyberinfrastructure Coordination Ecosystem: Services \& Support (ACCESS) program, supported by National Science Foundation grants \#2138259, \#2138286, \#2138307, \#2137603, and \#2138296. 

\section{Methods}\label{sec:methods}
\footnotesize
\paragraph{First-principles electron--phonon Hamiltonian in real space}\label{sec:hamiltonian} 
Modern solid-state density functional theory (DFT) and density functional perturbation theory (DFPT) have enabled first-principles electron--phonon Hamiltonians in reciprocal space~\cite{wannier,perturbo,epw,qe,wannier90}. To interface these Hamiltonians with our real-space dynamics simulations, we construct a fully real-space representation.

\paragraph{Electrons and phonons.} The electronic part is written in localized Wannier orbital basis,
\begin{equation}
\hat{H}_{\mathrm{el}} = \sum_{i,j, \mathbf R, \mathbf R_{\mathrm{e}}} t_{ij}(\mathbf R_{\mathrm{e}}) a_{i, \mathbf R}^\dagger a_{j, \mathbf R+\mathbf R_{\mathrm{e}}}\label{eq:real_He}
\end{equation}
where $t_{ij}(\mathbf R_{\mathrm{e}})$ is the hopping integral between the localized Wannier orbitals $i$ and $j$ whose cells are separated by vector $\mathbf R_{\mathrm{e}}$.
The Wannier orbitals are obtained from the unitary transformation $U_{\mathbf{k}}$ of the electronic Hamiltonian in reciprocal space $\hat{H}_{\mathrm{el},\mathbf{k}}$:
\begin{equation}
  t(\mathbf{R}_e)=\frac{1}{N}\sum_{\mathbf{k}} e^{-i \mathbf{k}\mathbf{R}_\mathrm e} U_{\mathbf{k}}^{\dagger} \hat{H}_{\mathrm{el},\mathbf{k}} U_{\mathbf{k}},
\end{equation}
\begin{equation}
  \hat{H}_{\mathrm{el}} = \sum_{m,\mathbf{k}} \varepsilon_{m}(\mathbf k) a_{m,\mathbf k}^\dagger a_{m,\mathbf{k}}
\end{equation}
where $N$ is the number of unit cells in the Born-von Kármán (BvK) supercell,
and $m,n$ are the Bloch band indices. The phonon dispersion is
\begin{equation}
H_{\mathrm{ph}}=\sum_{\nu,\mathbf q}\omega_\nu(\mathbf q)\,
b^\dagger_{\nu,\mathbf q}b_{\nu,\mathbf q},
\end{equation}
whose eigenmodes and frequencies are computed from diagonalizing the dynamical matrix,
\begin{equation}\label{eq:diag_Dq}
  D(\mathbf{q})\left|\mathbf{e}_\nu(\mathbf{q})\right\rangle=\omega_\nu(\mathbf{q})^2\left|\mathbf{e}_\nu(\mathbf{q})\right\rangle,\; \mathbf{m}_{a,\alpha,\nu}({\mathbf{q}}) = \frac{\mathbf{e}_{a,\alpha,\nu}(\mathbf{q})}{\sqrt{M_a}},\;
  \end{equation}
where $M_a$ is the mass of the atom $a$.
The interatomic force-constant (IFC) Hessian is computed from the DFPT calculations.
Phonons on a finer grid of $\mathbf{q}$-points are obtained by interpolating the IFC matrix $\Phi$,
\begin{equation}
  D_{\alpha\beta}^{ab}(\mathbf{q}) = \sum_{\mathbf{R}_{\mathrm{p}}} \Phi_{\alpha\beta}^{ab}(\mathbf{R}_{\mathrm{p}}) \mathrm{e}^{\mathrm{i} \mathbf{q} \cdot \mathbf{R}_{\mathrm{p}}} \cdot \frac{1}{\sqrt{M_a M_b}}
\end{equation}
where $a,b$ are the atom indices, and $\alpha,\beta$ are Cartesian indices.

\paragraph{Electron--phonon coupling.}
The general reciprocal-space expression for electron--phonon coupling (EPC) is written as
\begin{equation}\label{eq:ephreciprocal}
  H_{\mathrm{e-ph}} = \frac{1}{\sqrt{N}} \sum_{mn}\sum_{\mathbf k,\nu,\mathbf q} g_{mn}^\nu(\mathbf k,\mathbf q) a_{m,\mathbf k+\mathbf q}^\dagger a_{n,\mathbf k} (b_{\nu,\mathbf q} +b_{\nu,-\mathbf q}^\dagger)
\end{equation}
where $g_{mn}^\nu(\mathbf k,\mathbf q)$ is the electron--phonon coupling strength between the Bloch states $m$ and $n$ with wavevectors $\mathbf k$ and $\mathbf k+\mathbf q$, and the phonon mode $\nu$ with wavevector $\mathbf q$. We work with a fully real-space representation,
\begin{equation}
  H_{\mathrm{e-ph}} = \sum_{\substack{ij\nu\\\mathbf R_\mathrm e,\mathbf R_\mathrm p,\mathbf R}} g_{ij}^\nu(\mathbf R_{\mathrm e}, \mathbf R_{\mathrm p})a_{i,\mathbf R}^\dagger a_{j,\mathbf R+\mathbf R_{\mathrm e}} (b_{\nu,\mathbf R+\mathbf R_{\mathrm p}} + b_{\nu,\mathbf R+\mathbf R_{\mathrm p}}^\dagger)
\end{equation}
\begin{equation}
  (b_{\nu, \mathbf{R}+\mathbf{R}_{\mathrm{p}}}+b_{\nu, \mathbf{R}+\mathbf{R}_{\mathrm{p}}}^{\dagger}) = \frac{1}{\sqrt{N}} \sum_{\mathbf{q}} e^{\mathrm i \mathbf{q} \cdot (\mathbf{R}+\mathbf{R}_{\mathrm{p}})} (b_{\nu,\mathbf{q}} + b_{\nu,-\mathbf{q}}^\dagger)
\end{equation}
where $g_{ij}^\nu(\mathbf R_{\mathrm e}, \mathbf R_{\mathrm p})$ represents the coupling of the phonon mode $\nu$ moving at cell $\mathbf R_\mathrm p$ to the hopping between
the localized Wannier orbitals $i$ at origin and $j$ at cell $\mathbf R_\mathrm e$, which can be obtained by the 
following Fourier transformation,
\begin{equation}
  g_{ij}^\nu(\mathbf R_{\mathrm e},\mathbf R_{\mathrm p}) = \frac{1}{N}\sum_{\mathbf{q}} e^{-\mathrm{i} \mathbf{q} \cdot \mathbf R_{\mathrm p}} g_{ij}^\nu(\mathbf R_{\mathrm e}, \mathbf q).
\end{equation}
However, $g_{ij}^\nu(\mathbf R_{\mathrm e}, \mathbf q)$ is complex-valued and carries a gauge-dependent phase, 
which causes the direct Fourier transformation to yield delocalized electron--phonon couplings in real space $\mathbf R_{\mathrm p}$.
The random phase comes from the phonon mode $\mathbf e_{\nu, \alpha, \mathbf q}$,
\begin{equation}
  g_{ij}^\nu(\mathbf{R}_\mathrm{e}, \mathbf{q}) = \sum_{a \alpha,\mathbf R_\mathrm{p}} 
  \sqrt{\frac{\hbar}{2 \omega_\nu(\mathbf{q})}} \frac{\mathbf{e}_{a, \alpha, \nu}(\mathbf{q})}{\sqrt{M_a}} g_{i j}^{a \alpha}\left(\mathbf{R}_{\mathrm{e}}, \mathbf{R}_\mathrm{p}\right) e^{\mathrm{i} \mathbf{q} \cdot \mathbf{R}_\mathrm{p}} 
\end{equation}
where
\begin{align}
  g_{i j}^{a \alpha}\left(\mathbf{R}_\mathrm{e}, \mathbf{R}_\mathrm{p}\right) & =
  \frac{1}{N_{\mathbf{k}} N_{\mathbf{q}}} \sum_{m n \mathbf k \mathbf{q}} e^{-\mathrm i\left(\mathbf{k} \mathbf{R}_\mathrm e+\mathbf{q} \mathbf{R}_{\mathrm{p}}\right)} \\ \nonumber & 
   U_{i m}^{\dagger}\left(\mathbf{k}+\mathbf{q}\right) \Delta V_{m n, a \alpha}^S\left(\mathbf{k}, \mathbf{q}\right) U_{n j}\left(\mathbf{k}\right)
\end{align}
which characterizes the phonon modulation of the hopping integral between two localized Wannier functions 
(one at the origin and one at $\mathbf{R}_\mathrm{e}$), due to the displacement of atom $a$ at cell $\mathbf{R}_\mathrm p$ in direction $\alpha$.

\paragraph{Gauge-invariant maximally localized electron--phonon coupling.}

To make the electron--phonon coupling matrix $g_{ij}^\nu(\mathbf R_{\mathrm e}, \mathbf R_{\mathrm p})$ localized in real space, we introduce a gauge-invariant maximal-localization scheme,
\begin{equation}\label{eq:localization}
  \min _{\theta_\nu(\mathbf{q})}: \mathcal{L}=\sum_{i, j, \nu, \mathbf{R}_e} \sum_{\mathbf{R}_\mathrm p} \frac{\left|g_{i j}^\nu\left(\mathbf{R}_\mathrm e, \mathbf{R}_\mathrm p, \theta_\nu(\mathbf{q})\right)\right|^2}{\langle \omega_\nu\rangle_{\mathbf q}}(|\mathbf{R}_\mathrm p| - \delta_{\mathbf{R}_{\mathrm p}, \mathbf{R}_{\mathrm e}} - \delta_{\mathbf{R}_{\mathrm p},\mathbf{0}})
\end{equation}
where $\theta_\nu(\mathbf{q})$ is the gauge-dependent phase of the phonon mode $\nu$ with wavevector $\mathbf{q}$ to be optimized.
This localization scheme is designed to maximally localize electron--phonon coupling $g_{ij}^\nu(\mathbf R_{\mathrm e}, \mathbf R_{\mathrm p})$ in real space, with stronger weight assigned to configurations where the phonon displacement $\mathbf{R}_\mathrm{p}$ is closer to either the origin or the electronic hopping vector $\mathbf{R}_\mathrm{e}$ that directly couples the Wannier orbitals,
due to the following time-reversal symmetry,
\begin{equation}
  g_{ij}^\nu(\mathbf{R}_\mathrm{e}, \mathbf{R}_\mathrm{p}) = g_{ji}^\nu(-\mathbf{R}_\mathrm{e}, \mathbf{R}_\mathrm{p}-\mathbf{R}_\mathrm{e})
\end{equation}
With this gauge choice, compact electron--phonon coupling matrices are obtained (Supplementary Fig.~\ref{fig:localization}).

In a chosen phonon gauge, the branch-resolved real-space coupling 
$g_{ij}^{\nu}(\mathbf R_{\mathrm e},\mathbf R_{\mathrm p})$ is most naturally
interpreted as the coupling between the electronic matrix element 
$i(\mathbf 0)\!\to\! j(\mathbf R_{\mathrm e})$ and a \emph{localized real-space phonon basis state} 
(or phonon Wannier mode) of branch $\nu$ centered at $\mathbf R_{\mathrm p}$.
Here, the index $\mathbf R_{\mathrm p}$ labels the center of this localized phonon basis function, 
rather than the unique source cell of an atomic displacement. 
This is distinct from the Cartesian representation $g_{ij}^{a\alpha}(\mathbf R_{\mathrm e},\mathbf R_{\mathrm p})$, in which $\mathbf R_{\mathrm p}$ directly labels the cell containing the displaced atom.
This localized phonon branch label should be understood as a basis index rather than a literal phonon real-space coordinate. By design, the component denoted $\mathbf R_p=\mathbf{0}$ is the maximally localized branch with the largest weight near the electronic matrix element being modulated. It is this localized component that is shown in Fig.~\ref{fig:dntt}e for DNTT. The physical real-space content of the coupling can be obtained by expanding in the original real-space atomic indices with the optimized gauge.
Accordingly, the transformation from $g_{ij}^{a\alpha}$ to $g_{ij}^{\nu}$ mixes contributions from multiple source cells, so the branch-resolved coupling does not, in general, inherit the exact cell-by-cell support of the Cartesian coupling. As a result, a nonzero $g_{ij}^{\nu}(\mathbf R_{\mathrm e},\mathbf R_{\mathrm p})$ may arise even when all Cartesian couplings associated with atomic displacements in the same cell $\mathbf R_{\mathrm p}$ vanish, provided that the localized phonon basis state centered there has finite weight on neighboring cells that do contribute. In this representation, the phonon operators remain exact canonical bosonic operators, but the phonon Hamiltonian is no longer diagonal in $\mathbf R_{\mathrm p}$ (shown with an example in Supplementary Fig.~\ref{fig:omegedeltaR}); instead, the phonon dispersion appears as inter-cell couplings between these localized phonon basis states:
\begin{equation}
  H_{\mathrm{ph}}=\sum_\nu \sum_{\mathbf{R}, \Delta\mathbf{R}} \left(\frac{1}{N} \sum_{\mathbf{q}} \omega_\nu(\mathbf{q}) e^{i \mathbf{q} \cdot \Delta \mathbf{R}}\right)
  b_{\nu, \mathbf{R}}^{\dagger} b_{\nu, \mathbf{R}+\Delta\mathbf{R}}
  \label{eq:phonon_realspace}
\end{equation}

\paragraph{Particle-hole transformation.}\label{sec:pht}
To simulate hole transport, we apply a particle-hole transformation by 
interchanging the electronic creation and annihilation operators. 
This transformation necessitates inverting the signs of both the transfer integrals and 
the electron--phonon coupling matrix elements in the Hamiltonian 
to correctly describe the dynamics of positive charge carriers.

\paragraph{Computational details.}
Periodic DFT simulations 
were performed with Quantum ESPRESSO~\cite{qe}, using the PBE functional~\cite{pbe} with Grimme-D3 dispersion corrections~\cite{d3} and norm-conserving pseudopotentials (Pseudo Dojo~\cite{dojo}) to relax atomic positions and lattice parameters.
Wannier functions and transformation matrices were generated using Wannier90~\cite{wannier90}.
We calculated lattice dynamics and electron--phonon perturbation potentials via density functional perturbation theory (DFPT).
A $3\times 3\times 2$ $\mathbf k$-point grid was used for self-consistent field (SCF) calculations, followed by a $6\times 6\times 4$ non-self-consistent field (NSCF) calculation for Wannierization.
A $3\times 3\times 2$ $\mathbf q$-point grid was employed for phonon calculations.
Increasing the $\mathbf q$-point grid to $4\times 4 \times 2$ was used to test size convergence of the coarse-grid phonon calculations (Supplementary Fig.~\ref{fig:dntt_qgrid}).
Wannier interpolation and localization were performed on $\mathbf q$-point grids commensurate with the real-space simulation cells, with grids up to $18\times18\times1$ used for the largest systems. Grid-size convergence was verified for the transport observables.
The maximally localized electron--phonon coupling matrix is obtained by 
minimizing the localization functional (Eq.~\eqref{eq:localization}) using the L-BFGS-LS algorithm implemented in the \texttt{QCPBC} package.
All starting crystal structures were taken from experimental crystallographic data. DNTT was based on the X-ray structural analysis of Yamamoto and Takimiya~\cite{yamamoto2007facile}; pentacene was taken from COD 2012158; picene was taken from COD 5000182; C8-BTBT was taken from the CCDC-deposited structure CCDC 679293; and C8-DNTT was taken from the powder-X-ray-refined structure reported by Schweicher \textit{et al.}~\cite{schweicherChasingKillerPhonon2019}. Space-group symmetry was preserved during relaxation of both the cell parameters and atomic positions.



\paragraph{Mobility and optical conductivity with Green--Kubo formalism.}
Real-time propagation offers a general route to compute mobility and optical conductivity without invoking hopping or band mechanism assumptions.
While mobility can be extracted from the long-time limit of the mean squared displacement (MSD) of charge carriers via
\begin{equation}\label{eq:diffusion}
  \mu = \frac{eD(\tau\to\infty)}{k_B T}, \quad D(t) = \frac{1}{2}\frac{d}{d\tau}\langle\Delta x(\tau)^2\rangle,
\end{equation}
this approach requires large supercells and long simulation times to reach the thermodynamic diffusive limit.
Furthermore, long-time propagation in mixed quantum--classical dynamics is often compromised by the violation of detailed balance and artificial overheating.
Instead, we employ the Green-Kubo formalism, which calculates the current--current correlation function and optical conductivity tensor in $\alpha,\beta$ direction:
\begin{equation}\label{eq:Ct}
C_{\alpha\beta}(\tau) = \operatorname{Tr}(J_\alpha(\tau) J_\beta(0)\rho_T), \quad \mu_{\alpha\beta} = \frac{\int_0^\infty C_{\alpha\beta}(\tau) d\tau}{k_B T}
\end{equation}
\begin{equation}
  \frac{\sigma_{\alpha\beta}(\omega)}{n e_0}=\frac{1-e^{-\omega / k_B T}}{\omega} \int_0^{\infty} C_{\alpha\beta}(t) e^{i \omega t} \mathrm{~d} t
\end{equation}
Because $C(\tau)$ naturally decays to zero, the mobility integral converges at short to intermediate times where the dynamics remain accurate, rather than requiring the long-time limit where mixed quantum--classical methods suffer from detailed-balance violations and artificial energy drift, and even numerically exact methods face growing computational cost. This makes the Green--Kubo route substantially more robust for extracting quantitative transport coefficients.
For the largest systems, however, propagating trajectories until $C(\tau)$ decays to zero is computationally demanding and increases exposure to detailed-balance violations and artificial energy drift in mixed quantum--classical dynamics. We instead combine the explicitly calculated short-time current response with memory-kernel reconstruction, reducing both the required propagation time and sensitivity to these long-time artifacts.
Finite-size convergence was checked using lattices containing up to \(18\times18\) unit cells (648 molecules), corresponding to more than 160,000 phonon degrees of freedom for the largest system.

\paragraph{Memory-kernel reconstruction.}
To extend the explicitly propagated current correlation function to long times, we extract the Mori memory kernel \(M(t)\), which characterizes non-Markovian effects, from
\begin{equation}
\frac{dC(t)}{dt}
=
-\int_0^t M(t-\tau)C(\tau)\,d\tau .
\end{equation}
The random-force contribution vanishes after projection onto the current-autocorrelation function by orthogonality, so that \(M(t)\) is determined from \(C(t)\).
When the memory kernel decays within the explicit propagation window, the correlation dynamics can be represented using a finite memory time.
The converged short-time memory kernel is then used in the above Volterra equation to reconstruct the long-time current correlation function and evaluate the Green--Kubo integral.

\paragraph{Finite temperature time-dependent matrix product states.}
To obtain exact reference results for the one-dimensional benchmarking cases, we use finite-temperature time-dependent matrix product states (MPS):
\begin{equation}
    |\Psi\rangle=\sum_{\{a\},\{\sigma\}} A_{a_1}^{\sigma_1} A_{a_1 a_2}^{\sigma_2} \cdots A_{a_{N-1}}^{\sigma_N}\left|\sigma_1 \sigma_2 \cdots \sigma_N\right\rangle
\end{equation}
The equilibrium state is constructed via thermal field dynamics (purification), where the thermal density matrix is expressed as a partial trace over a pure state $|\Psi_\beta\rangle$ defined in an enlarged Hilbert space. This state is generated through imaginary-time evolution from a maximally entangled initial state,
\begin{equation}
    \frac{e^{-\beta H}}{Z}=\frac{\operatorname{Tr}_Q\left|\Psi_\beta\right\rangle\left\langle\Psi_\beta\right|}{\operatorname{Tr}_{P Q}\left|\Psi_\beta\right\rangle\left\langle\Psi_\beta\right|}
\end{equation}
Real-time evolution is then applied to compute Eq.~\eqref{eq:Ct} without approximating the phonon effects in the benchmark models.
The current operator is represented as a matrix product operator (MPO). Time evolution is performed using a variational principle-based projector splitting method, offering both accuracy and efficiency when combined with GPU acceleration.

\paragraph{Mixed quantum-classical dynamics.}

To simulate the dynamics for the first-principles Hamiltonian,
by choosing a cutoff frequency $\omega_c=\max(2k_BT, |t|_{\mathrm{max}})$ to separate the fast and slow phonons,
we approximate the current-current correlation function (Eq.~\eqref{eq:Ct}) with mixed treatment of phonons.
\begin{align}
  C(\tau) & = \int d \mathbf{X}_{\mathrm{slow}} \int d \mathbf{P}_{\mathrm{slow}} 
  \mathcal{P}\left(\mathbf{X}_{\mathrm{slow}}, \mathbf{P}_{\mathrm{slow}}\right) \\ \nonumber
  & \times \operatorname{Tr}\left[U_\mathrm{fast}(0, \tau) J(\tau) 
  U_\mathrm{fast}(\tau, 0) J e^{-\beta H_\mathrm{fast}(0)}\right] / Z_\mathrm{fast}
\end{align}
where the low-frequency phonons are sampled from the Boltzmann or Wigner distribution and averaged over trajectories.
The electronic degree of freedom and the fast phonons (rapidly oscillating phonons) are both treated quantum mechanically,
\begin{equation}
  U_{\mathrm{fast}}(\tau, 0)=T \exp \left[-i \int_0^\tau d \tau^{\prime} H_\mathrm{fast}\left(\tau^{\prime}\right)\right]
\end{equation}
The high-frequency phonons are coupled mainly to the onsite energies of the electronic degrees of freedom,
and a variational polaron transformation~\cite{chengUnifiedTheoryChargecarrier2008,wangVariationalPolaronTransformation2020} is used to trace out the high-frequency phonons analytically,
\begin{equation}
  \hat{G}=\sum_{i,R,\nu} f_\nu\left(b_{\nu,\mathbf{R}}^{\dagger}-b_{\nu,\mathbf{R}}\right) a_{i,\mathbf{R}}^{\dagger} a_{i,\mathbf{R}},\quad \tilde{O}=e^{\hat{G}}Oe^{-\hat{G}}
\end{equation}
The variational parameters $\{f_\nu\}$ are determined by minimizing the polaronic free energy $A_\beta$, 
\begin{equation}
  A_\beta = -\beta^{-1} \ln \operatorname{Tr} e^{-\beta \tilde{H}_\mathrm{fast}}
\end{equation}
which allows the fast phonons to partially renormalize the electronic hopping parameters in the presence of slow phonon fluctuations, for $\mathbf{R}$ at the origin,
\begin{align}
  \tilde{t}_{ij}(\mathbf{R}_{\mathrm{e}}, \tau) & =\left(t_{ij}(\mathbf{R}_{\mathrm{e}})+\sum_{\nu,\mathbf{R}_{\mathrm{p}}} g_{ij}^\nu(\mathbf{R}_{\mathrm{e}}, \mathbf{R}_{\mathrm{p}}) X^\nu_{\mathbf{R}_{\mathrm{p}}}(\tau)\right) \\ \nonumber
  & \times \exp \left[-\sum_\nu f_{\nu}^2 \operatorname{coth}\left(\beta \omega_\nu / 2\right)\right]
\end{align}
The electronic tight binding Hamiltonian, both the bare electronic hopping parameters
and the slow phonon displacements are partially dressed by the fast phonons and lead to 
bandwidth narrowing. For a single trajectory, the corresponding current correlation function is
\begin{equation}
  C_{\mathrm{itraj}}(\tau)=\operatorname{Tr}\left[\tilde{J}_{\mathrm{itraj}}(\tau) \tilde{U}_{\mathrm{fast,itraj}}(\tau, 0) \tilde{J}_{\mathrm{itraj}}(0) \tilde{\rho}_{\mathrm{itraj}}(\tau) \tilde{U}_{\mathrm{fast,itraj}}^{\dagger}(\tau, 0)\right]
\end{equation}
which will be numerically evaluated by optimized tensor contractions that make use of the sparsity of the Hamiltonian.

\paragraph{Tight-binding analysis}
To analyze the impact of lattice fluctuations, we performed a statistical study of the tight-binding model by sampling 10,000 nuclear configurations 
derived from low-frequency phonon modes.
For each trajectory, we extracted the effective Hamiltonian and diagonalized it to obtain the 
instantaneous eigenstates, which were then ensemble-averaged for static analysis. 
We assessed the spatial delocalization of the carrier by computing a density-matrix coherence measure and the participation ratio (PR)~\cite{thouless1974electrons},
\begin{equation}
  L_{\mathrm{loc}} =\sum_{i\ge j} |\rho_{ij}|,\quad 
  \operatorname{PR}= \left \langle \frac{1}{\sum_i |p_i|^4} \right \rangle_{N_{\mathrm{itraj}}}
\end{equation}
For a normalized density matrix, $L_{\mathrm{loc}}=1$ for a fully incoherent diagonal state with uniform populations, $\rho_{ij}=\delta_{ij}/N$, whereas a fully coherent state uniformly delocalized over $N$ molecules, $\rho_{ij}=1/N$, gives $L_{\mathrm{loc}}=(N+1)/2$. The delocalization sizes shown in Fig.~\ref{fig:pentacene_DNTT}d are normalized using the simulated lattice so that values 
correspond to equivalent-area delocalization diameter and
can be compared across the materials studied here.
The reciprocal fourth moment is the participation ratio introduced in localization theory~\cite{thouless1974electrons}; with this convention, larger PR values indicate more spatially extended states.

The agreement between this static tight-binding analysis and the transport observables from real-time dynamics provides two complementary views of the same localization hierarchy.
For the real-space coherence maps in Fig.~\ref{fig:pentacene_DNTT}d, we constructed the thermally averaged single-particle density matrix from the sampled Hamiltonians and plotted the coherence relative to a central molecular site, \(\langle a_{\mathrm{center}}^\dagger a_j\rangle_{\mathrm{eq}}\). The distributions of \(L_{\mathrm{loc}}\) provide the corresponding ensemble-level localization statistics.

Following the Green-Kubo expression, the energy-resolved optical-conductivity weight can be defined as~\cite{li2014optical},
\begin{align}\label{eq:Xi}
\Xi(E_i,E_f)
& =
\langle
\sum_{m,n;\,E_n>E_m}
\frac{p_m-p_n}{E_n-E_m}
\left(
|\langle m|J_x|n\rangle|^2+
|\langle m|J_y|n\rangle|^2
\right) \\ \nonumber 
& \times \delta(E_i-E_m)\delta(E_f-E_n)
\rangle_s
\end{align}
Here \(E_i\) and \(E_f\) are the initial and final hole-state energies, and
\(\langle\cdots\rangle_s\) denotes the average over sampled nuclear configurations.
The averaged optical conductivity at photon energy \(\hbar\omega\) can be approximated by projecting
\(\Xi(E_i,E_f)\) along constant transition-energy diagonals,
\begin{equation}
\sigma(\omega) \propto
\int dE_i\,dE_f\,
\Xi(E_i,E_f)\,
\delta(E_f-E_i-\hbar\omega).
\end{equation}
Thus \(\Xi(E_i,E_f)\) separates the conductivity into contributions from thermally
occupied initial states, available final states, and current matrix elements.

\paragraph*{Data and code availability:}
Data supporting the conclusions of this study are provided in the paper, Supplementary Materials, and the repository described below.
The processed data and plotting scripts supporting the findings of this study have been deposited at \url{https://github.com/JoonhoLee-Group/first-principles-transport-data}.
The custom first-principles modelling and quantum-dynamics software developed in this work, \textsc{PyEPH}, is available at \url{https://github.com/jiangtong1000/PyEPH}.
Versioned archival releases of the supporting data and code, including input files and workflow settings, will be provided with persistent DOIs upon publication.

\bibliography{main.bib}

\clearpage
\newpage
\onecolumngrid
\linespread{1.3}\selectfont                                                                                                                

\renewcommand{\thefigure}{S\arabic{figure}}
\renewcommand{\thetable}{S\arabic{table}}
\renewcommand{\theequation}{S\arabic{equation}}
\renewcommand{\thepage}{S\arabic{page}}
\renewcommand{\thesection}{S\arabic{section}}
\renewcommand{\thesubsection}{\arabic{subsection}}
\setcounter{figure}{0}
\setcounter{table}{0}
\setcounter{equation}{0}
\setcounter{section}{0}
\setcounter{page}{1} 
\renewcommand{\theHfigure}{S\arabic{figure}}
\renewcommand{\theHtable}{S\arabic{table}}
\renewcommand{\theHequation}{S\arabic{equation}}
\renewcommand{\theHsection}{S\arabic{section}}
\renewcommand{\theHsubsection}{S\arabic{subsection}} 

\setcounter{secnumdepth}{4}


\begin{center}
{\Large \textbf{Supplementary Materials for}}\\[0.5em]
{\large \textbf{First-Principles Origins of Charge Transport in Molecular Semiconductors}}\\[0.5em]
{Tong Jiang, Joonho Lee}\\
{Department of Chemistry and Chemical Biology, Harvard University\\12 Oxford St, Cambridge, Massachusetts 02138, USA}
\end{center}

\section{Supplementary Notes and Figures}

\paragraph*{Methodological overview.}
The central methodological task in this work is to connect first-principles solid-state electron--phonon Hamiltonians to quantitative finite-temperature transport without imposing a band, hopping, or transient-localization/delocalization model in advance. This requires three ingredients: construction of material-specific solid-state Hamiltonians, a gauge-invariant localization procedure that makes these Hamiltonians usable for real-space dynamics, and a non-perturbative Green--Kubo dynamics solver validated against numerically exact or high-level model benchmarks.
\begin{figure}[htbp]
  \centering
  \includegraphics[width=0.8\textwidth]{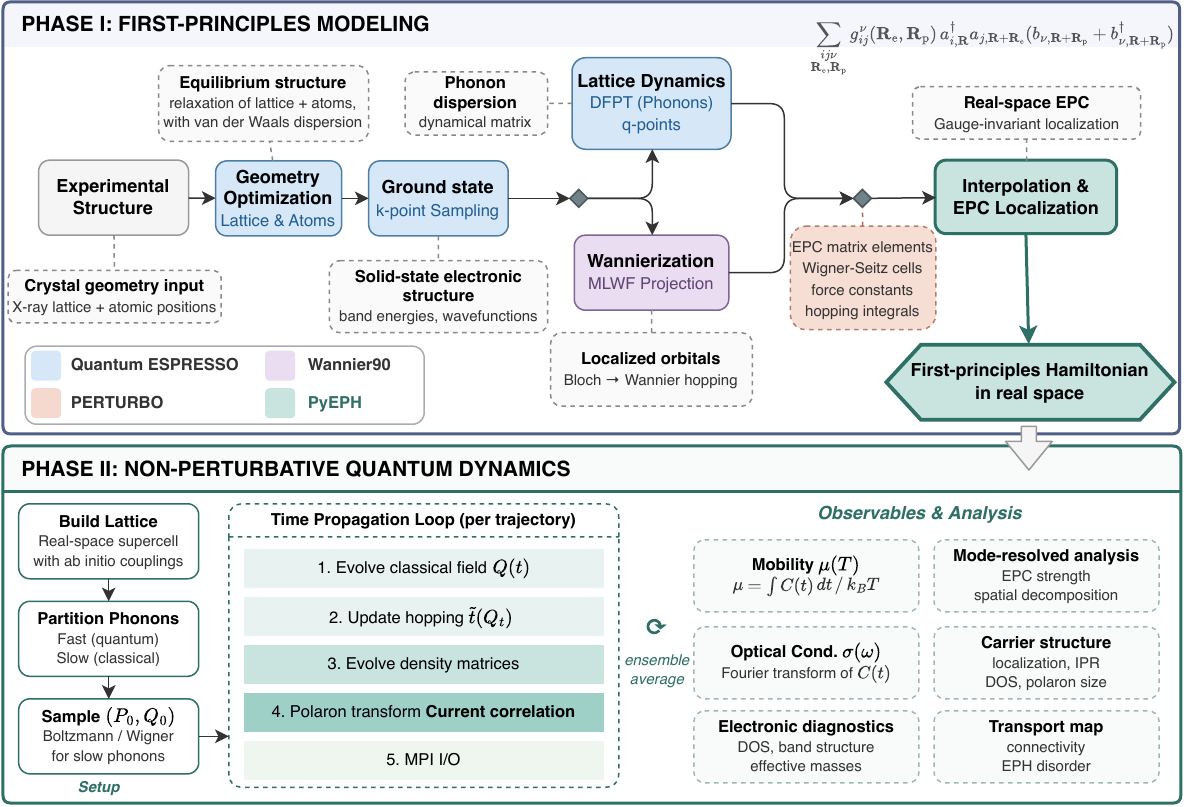}
  \caption{\textbf{Detailed computational workflow.}
  \textit{Phase~I}: First-principles modeling, from experimental crystal structure through DFT/DFPT electronic structure and lattice dynamics, Wannierization, and electron--phonon coupling localization to a real-space Hamiltonian.
  \textit{Phase~II}: Non-perturbative quantum dynamics, comprising lattice construction, phonon partitioning, thermal sampling, time propagation (classical field evolution, hopping updates, polaron transformation, current correlation), and analysis of transport observables.}
  \label{fig:workflow}
\end{figure}

This is an end-to-end workflow that combines mature first-principles tools with our new infrastructure for real-space Hamiltonian construction and dynamics.
Standard solid-state components, including periodic density functional theory (DFT) and density functional perturbation theory (DFPT) calculations, Wannierization, and reciprocal-space electron--phonon interpolation, are performed using established packages such as Quantum ESPRESSO, Wannier90, and Perturbo~\citeS{qe,wannier90,perturbo}.
To make these Hamiltonians usable for real-space quantum dynamics, we developed the \texttt{PyEPH} (https://github.com/jiangtong1000/PyEPH) workflow, which interfaces with reciprocal-space electron--phonon data, performs interpolation and gauge-invariant localization, and constructs compact real-space Hamiltonians retaining both electronic locality and solid-state phonon information.

On top of these localized Hamiltonians, \texttt{PyEPH} implements the non-perturbative Green--Kubo dynamics used to compute mobility and optical conductivity. The solver treats slow lattice disorder and fast phonon dressing within a unified framework, using classical sampling for low-frequency modes and variational polaron dressing for higher-frequency modes. Because this mixed quantum--classical strategy is approximate, we also carried out systematic benchmarks against numerically exact or high-level reference calculations on lattice Hamiltonians before applying the method to first-principles molecular-crystal Hamiltonians. Where reference data were unavailable, we generated finite-temperature tensor-network benchmarks using \texttt{Renormalizer} (https://github.com/shuaigroup/Renormalizer).
Figure~\ref{fig:workflow} summarizes the end-to-end computational pipeline, from experimental crystal structure through first-principles electronic structure and lattice dynamics to nonperturbative quantum dynamics and transport observables.
Methods and formalisms are detailed in the Methods section.

\paragraph*{Benchmarking lattice models.}
To validate the mixed quantum--classical dynamics employed in this work, we benchmark across a hierarchy of electron--phonon models, summarized in main-text Fig.~\ref{fig:dntt}a.
These benchmarks include a one-dimensional SSH/Peierls model representing intermolecular coupling, a one-dimensional Holstein model representing intramolecular coupling, a two-dimensional Peierls model on a herringbone lattice with representative material parameters, and a multimode rubrene model combining nine intramolecular Holstein modes and one intermolecular Peierls mode at room temperature.
A closely related variational mixed quantum--classical treatment was introduced by Fetherolf \textit{et al.} for simple 1D Holstein--Peierls transport models~\citeS{fetherolf2020unification}. Here we go beyond that initial demonstration by carrying out direct head-to-head benchmarks against numerically exact finite-temperature matrix product state (MPS) simulations for 1D Holstein and Holstein--Peierls models, including a large multimode Holstein--Peierls case with nine Holstein modes and one Peierls mode that already pushes against the practical limit of current tensor-network calculations.
For the 1D SSH model we also compare with hierarchical equations of motion (HEOM) results~\citeS{jankovic2025charge1}. For the 2D SSH model on a herringbone lattice, where exact real-time benchmarks remain prohibitive, we compare against hybrid Monte Carlo (HMC) simulations~\citeS{PhysRevApplied.22.L031004}, which employ analytic continuation but still represent the best available reference. 
In the benchmarking setting, $t$ denotes the electronic hopping integral, $\omega$ the phonon frequency, $\lambda$ the Holstein reorganization energy for local intramolecular coupling, $g_{\mathrm P}$ the Peierls coupling strength for phonon-induced modulation of intermolecular hopping~\citeS{jankovic2025charge1}, and $\Delta t$ the root-mean-square intermolecular hopping fluctuation generated by low-frequency dynamic disorder~\cite{li2021general}.
The agreement across all regimes establishes the reliability of the solver for the parameter space relevant to the materials studied in the main text. Why this mixed quantum--classical strategy performs well in these regimes, despite its known limitations, is discussed further in Section~\ref{sec:connections}. Methodological details of both the mixed quantum--classical dynamics and the finite-temperature MPS simulations are provided in the Methods section. 

\begin{figure}[htbp]
  \centering
  \includegraphics[width=0.8\textwidth]{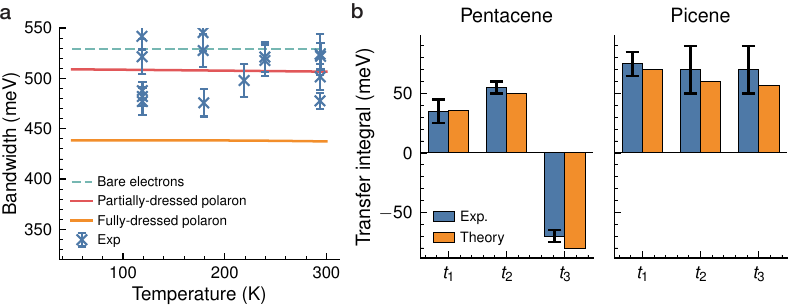}
  \caption{\textbf{Validation of first-principles electronic structure against photoemission experiments.}
  \textbf{(a)} HOMO bandwidth of pentacene: bare electron band, fully dressed (Lang--Firsov), and partially dressed (variational polaron transformation, this work), compared with ARPES measurements~\protect\citeS{satoConductionBandStructure2022}.
  \textbf{(b)} Electronic transfer integrals ($t_1$, $t_2$, $t_3$) for pentacene and picene from Wannier-localized orbitals (this work) and ARPES measurements~\protect\citeS{neef2024frontier}.
  }
  \label{fig:transfer_integrals}
\end{figure}
\begin{figure}[htbp]
  \centering
  \includegraphics[width=0.95\textwidth]{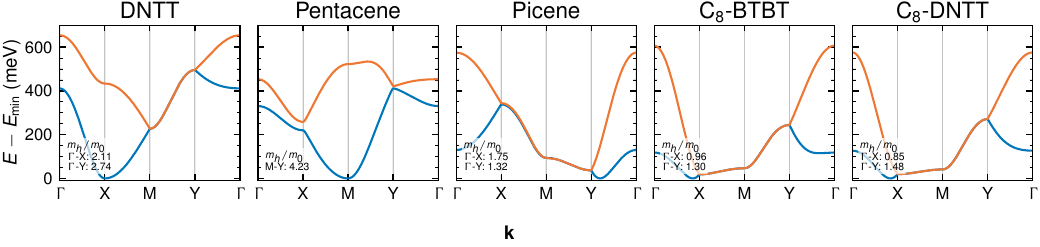}
  \caption{\textbf{Valence band dispersions of benchmark molecular semiconductors.}
  Band energies are referenced to the minimum energy in each panel.
  The two frontier hole bands are shown along the in-plane high-symmetry path for DNTT, pentacene, picene, C8-BTBT, and C8-DNTT.
  Insets report the hole effective masses in units of the free-electron mass.}
  \label{fig:bands}
\end{figure}

\paragraph*{Validation of first-principles electronic structure.}
As an independent check of the electronic Hamiltonian, we validate both the bandwidth and transfer integrals against photoemission experiments.
Figure~\ref{fig:bands} shows the corresponding first-principles frontier-band dispersions for the benchmark molecular semiconductors.
The band curvatures provide a direct electronic-structure view of the connectivity axis discussed in the main text: picene, C8-BTBT, and C8-DNTT show comparatively light in-plane hole masses, whereas DNTT and especially pentacene have heavier and more anisotropic band edges.
These dispersions therefore complement the transfer-integral analysis by showing how the localized Hamiltonian translates into the band-edge effective masses that enter the coherent limit of transport.
Figure~\ref{fig:transfer_integrals}a compares the computed HOMO bandwidth for pentacene with ARPES measurements~\citeS{satoConductionBandStructure2022}. The bare electronic bandwidth overestimates the experimental value, while a full Lang--Firsov polaron dressing underestimates it; the variational polaron transformation employed in this work yields a partially dressed bandwidth in quantitative agreement with experiment, confirming that the balance between slow and fast phonon dressing is correctly captured.
Figure~\ref{fig:transfer_integrals}b compares the three symmetry-distinct transfer integrals ($t_1$, $t_2$, $t_3$) for pentacene and picene with ARPES-extracted values~\citeS{neef2024frontier}, showing excellent agreement for both materials. Together, these comparisons validate the electronic structure that underpins all subsequent transport simulations.

\begin{figure}[htbp]
  \centering
  \includegraphics[width=\textwidth]{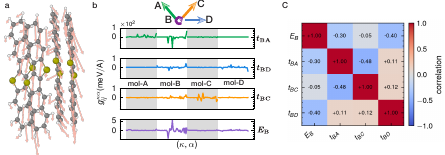}
  \caption{\textbf{(a)} Out-of-plane view of the acoustic phonon mode in DNTT.
  The same mode shown in Fig.~\ref{fig:dntt}e, viewed along the out-of-plane direction. 
  The displacement pattern has predominantly out-of-plane acoustic character at finite $\mathbf q$, producing relative molecular displacements rather than a uniform $\Gamma$-point translation, which is absent from isolated-molecule or dimer calculations.
  \textbf{(b)} Electron--phonon couplings projected onto the Cartesian atomic motions. The indices $\kappa$ and $\alpha$ denote the atom index and Cartesian direction, respectively. Hopping derivatives are concentrated mainly on the two molecules connected by the hopping, whereas onsite-energy derivatives are concentrated mainly on the molecule hosting the charge.
    \textbf{(c)} The Pearson correlation matrix among electronic-parameter fluctuations generated by the lowest phonon branch.
  } 
  \label{fig:acoustic}
\end{figure}

\paragraph*{The acoustic phonon mode in DNTT.}
Figure~\ref{fig:acoustic}a shows the lowest-frequency phonon mode ($\nu{=}1$) in DNTT from an out-of-plane perspective, complementing the in-plane view in Fig.~\ref{fig:dntt}e. From this viewpoint, the four molecules in the unit cell undergo finite-$\mathbf q$ acoustic distortions. This type of collective motion is an intrinsically solid-state vibration that arises from the periodic lattice environment and is absent from monomer or dimer calculations widely used to parameterize electron--phonon couplings for studying charge transport in organic semiconductors. That the transport-limiting mode in DNTT possesses this acoustic character underscores the necessity of a full solid-state treatment of lattice dynamics for quantitative transport predictions.

\begin{figure}[htbp]
  \centering
  \includegraphics[width=0.65\textwidth]{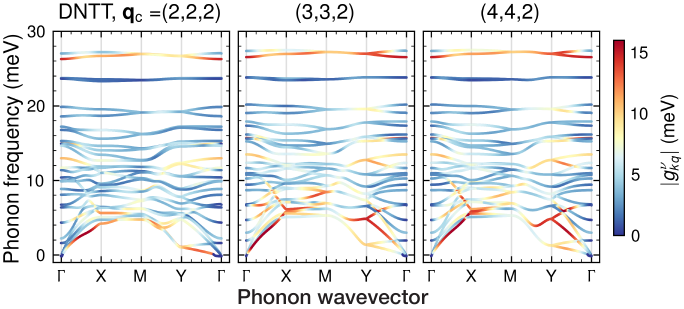}
  \caption{\textbf{Coarse-grid convergence of the DNTT phonon dispersion and electron--phonon coupling.}
  Phonon dispersions of DNTT overlaid with mode-resolved electron--phonon coupling strengths for coarse $\mathbf q$-point grids $\mathbf q_c=(2,2,2)$, $(3,3,2)$, and $(4,4,2)$. The strongest coupling features remain concentrated on the same low-frequency branches across all three samplings.}
  \label{fig:dntt_qgrid}
\end{figure}

\paragraph*{Coarse-grid convergence in DNTT.}
Figure~\ref{fig:dntt_qgrid} compares the DNTT phonon dispersions and mode-resolved electron--phonon coupling strengths obtained from coarse $\mathbf q$-point grids of $2\times 2\times 2$, $3\times 3\times 2$, and $4\times 4\times 2$. The overall dispersion and, more importantly, the concentration of the strongest coupling on the same low-frequency branches are already well captured at $3\times 3\times 2$.

\begin{figure}[htbp]
  \centering
  \includegraphics[width=0.8\textwidth]{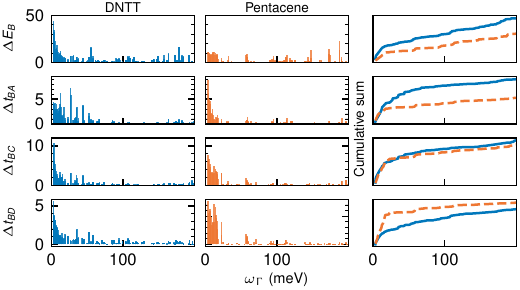}
  \caption{\textbf{Channel-resolved disorder comparison between DNTT and pentacene.}
Mode-resolved fluctuation amplitudes of the onsite energies and hopping integrals for DNTT and pentacene. 
Frequencies are represented by the corresponding $\Gamma$-point phonon frequencies for visualization. DNTT exhibits substantially stronger low-frequency disorder, especially in the onsite-energy channel, whereas pentacene has weaker coupling that is shifted to higher frequencies.}
  \label{fig:dntt_pentacene_disorder}
\end{figure}

\begin{figure}[htbp]
  \centering
  \includegraphics[width=0.5\textwidth]{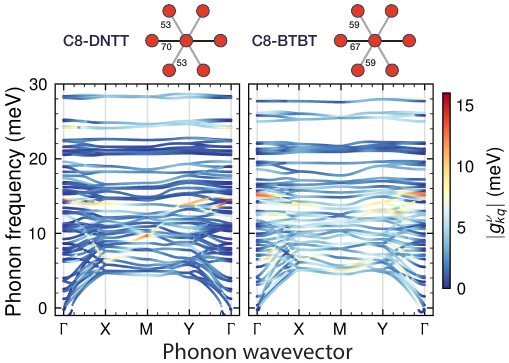}
  \caption{\textbf{Comparison of C8-DNTT and C8-BTBT.}
  Both materials are alkyl-modified $\pi$-conjugated cores with similar transfer integral networks. However, C8-DNTT exhibits weaker electron--phonon coupling than C8-BTBT, as revealed by the mode-resolved coupling strengths overlaid on the phonon dispersions. This weaker coupling underlies the higher intrinsic mobility of C8-DNTT (Fig.~\ref{fig:dntt}b).
  }
  \label{fig:dntt_btbt}
\end{figure}

\paragraph*{Comparison of C8-DNTT and C8-BTBT.}
As discussed in the main text, our simulations predict that C8-DNTT possesses a higher intrinsic mobility than C8-BTBT (Fig.~\ref{fig:dntt}b). Figure~\ref{fig:dntt_btbt} provides the microscopic basis: despite similar transfer-integral networks, C8-DNTT's electron--phonon coupling is systematically weaker, consistent with its higher mobility.

\begin{figure*}
    \includegraphics[width=0.95\textwidth]{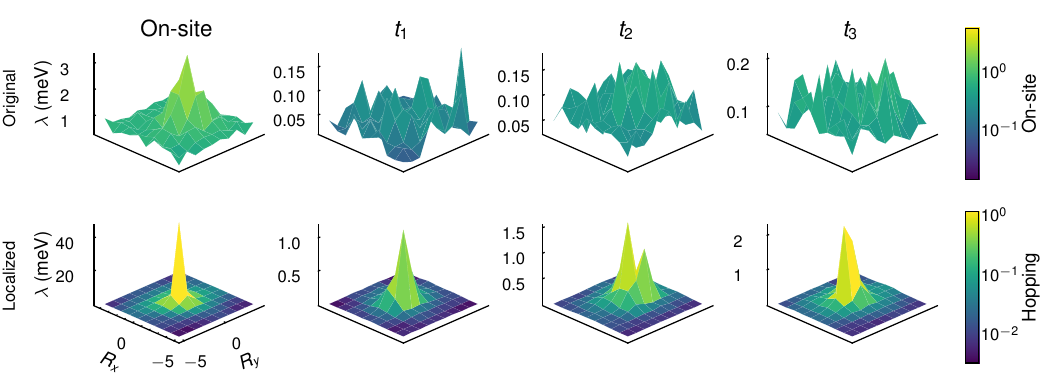}
\caption{\textbf{Gauge-invariant localization of real-space electron--phonon couplings.} Pentacene as an example: representative accumulated reorganization energy $\lambda$ is shown for the onsite energy and the three hopping channels $t_1$, $t_2$, and $t_3$. The top row shows the electron--phonon couplings obtained by direct Fourier transformation in an arbitrary phonon gauge, where gauge-dependent phonon phases produce delocalized real-space weight over $\mathbf R_{\mathrm p}=(R_x,R_y)$.
The bottom row shows the same quantities after optimizing the phonon phases with the localization functional in Eq.~\eqref{eq:localization}, which concentrates the coupling near the electronic matrix element being modulated.
}
    \label{fig:localization}
\end{figure*}

\paragraph*{Localization of real-space electron--phonon couplings.}
The reciprocal-space electron--phonon matrix elements carry gauge-dependent phonon phases, so a direct Fourier transform can produce artificially delocalized real-space couplings. Supplementary Fig.~\ref{fig:localization} illustrates this issue for pentacene and shows that the gauge optimization in Eq.~\eqref{eq:localization} concentrates the coupling near the electronic matrix element being modulated. This localized representation provides the compact real-space Hamiltonian used in the transport simulations.

\begin{figure}[htbp]
  \centering
  \includegraphics[width=\textwidth]{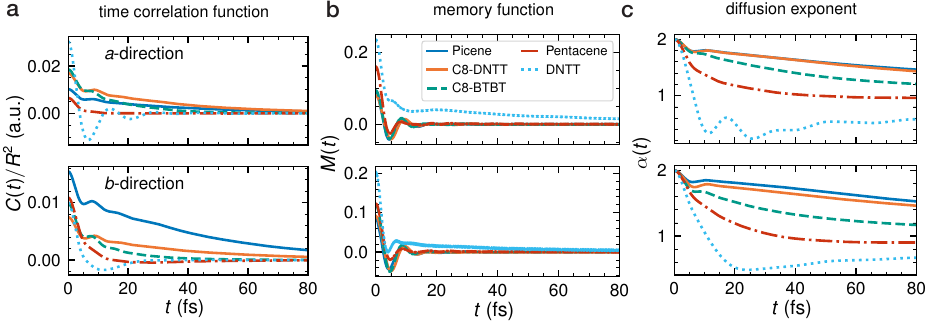}
  \caption{\textbf{Real-time transport dynamics of benchmark molecular semiconductors at room temperature.}
  \textbf{(a)} Current--current correlation functions $C(t)$. The decay timescale and oscillatory structure reflect the interplay between electronic coherence and phonon-induced decoherence in each material.
  \textbf{(b)} Memory function extracted with the Mori formalism, characterizing non-Markovian effects. DNTT exhibits a notably long-lived memory due to its strong coupling to low-frequency collective modes.
  \textbf{(c)} Time-dependent diffusion exponent \(\alpha(t)\), where \(\alpha=2\) indicates ballistic transport, \(\alpha>1\) superdiffusion, \(\alpha=1\) normal diffusion, and \(\alpha<1\) subdiffusion.
  }
  \label{fig:tcf}
\end{figure}

\paragraph*{Real-time dynamics.}
Figure~\ref{fig:tcf} presents the calculated current--current correlation functions, corresponding Mori memory kernels, and time-dependent diffusion exponents for the molecular semiconductors at room temperature.
The memory-kernel formulation and long-time reconstruction procedure are described in Methods.

To characterize the evolution from short-time ballistic motion toward longer-time transport regimes, we define the finite-time diffusion coefficient as the running integral of the current correlation function,
\begin{equation}
D(t)=\int_0^t C(t')\,dt'=\frac{1}{2}\frac{d}{dt}\langle \Delta x(t)^2\rangle .
\end{equation}
This quantity is exact at all times; the Einstein relation $\mu = eD/k_BT$ applies once $D(t)$ converges to a plateau, corresponding to the onset of normal diffusion ($\alpha = 1$).
We further define the time-dependent diffusion exponent $\alpha(t)$ from the mean squared displacement (MSD) as
\begin{equation}
\alpha(t)=\frac{d\ln \langle \Delta x(t)^2\rangle}{d\ln t},
\end{equation}
which distinguishes superdiffusive, diffusive, and subdiffusive transport regimes.
As shown in Fig.~\ref{fig:tcf}c, DNTT exhibits prolonged subdiffusive behavior ($\alpha < 1$) within the simulation window, consistent with its strong transient localization by low-frequency acoustic modes. Pentacene, by contrast, converges quickly and monotonically from ballistic ($\alpha = 2$) to diffusive ($\alpha = 1$) transport. The high-mobility materials C8-DNTT, picene, and C8-BTBT sustain superdiffusive exponents ($\alpha > 1$) over extended timescales, reflecting their more coherent carrier dynamics and consistent with their higher mobilities. 
The same current--current correlation functions determine the mobility and optical conductivity reported in the main text. Their contrasting decay profiles reflect differences in electron--phonon coupling: DNTT's long-lived, slowly decaying memory traces to its strong coupling to low-frequency acoustic modes (Fig.~\ref{fig:dntt}), whereas pentacene, picene, and C8-BTBT show more rapid decay consistent with their weaker couplings.

\begin{figure}
    \centering
    \includegraphics[width=0.35\linewidth]{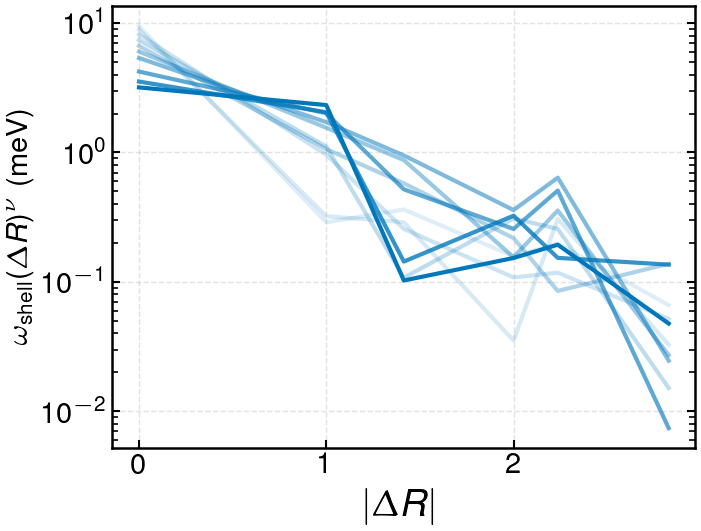}
\caption{\textbf{Nonlocality of the real-space phonon basis.}
Shell-resolved magnitudes of the real-space phonon Hamiltonian matrix elements
$\omega_{\mathrm{shell}}^{\nu}(\Delta \mathbf R)$ obtained by Fourier transforming the phonon dispersion $\omega_\nu(\mathbf q)$ of DNTT, as in Eq.~(\ref{eq:phonon_realspace}). Each curve corresponds to a phonon branch $\nu$.}
    \label{fig:omegedeltaR}
\end{figure}

\paragraph{Connections to alternative dynamics methods}\label{sec:connections}
A key design principle of this work is that the first-principles Hamiltonian is constructed independently of the quantum dynamics solver, and can serve as input to other advanced methods for electron--phonon coupled systems. 
Here, we briefly situate our solver choice within the broader landscape of available approaches.

Most existing transport predictions with first-principles flavors employ perturbative theories (Boltzmann transport, bubble approximations) applied to the reciprocal-space electron--phonon coupling matrix~\citeS{perturbo,epw,changIntermediatePolaronicCharge2022,changBandlikeChargeTransport2024}. While chemically specific, these assume weak coupling or well-defined quasiparticles and cannot capture the nonperturbative effects that govern transport in the intermediate regime.

Among nonperturbative approaches, trajectory surface hopping methods, including fewest-switches surface hopping and its decoherence-corrected variants, have been widely applied to charge transport in organic semiconductors~\citeS{wang2013flexible} with lattice model descriptions. Fragment orbital-based implementations~\citeS{giannini2019quantum} derive the electronic Hamiltonian from isolated molecular dimers or classical force-field-modulated fragments, enabling large-scale simulations but approximating the phonons. Mean-field Ehrenfest dynamics~\citeS{PhysRevLett.96.086601,wangMultiscaleStudyCharge2010} avoids stochastic hopping but suffers from incorrect branching and the absence of detailed balance. Most of these nonperturbative methods extract mobility from the long-time diffusive limit of the mean squared displacement via the Einstein relation, which requires propagating trajectories to times where mixed quantum--classical approximations generically suffer from violations of detailed balance and artificial energy drift. Related limitations were identified earlier by Berne and co-workers in mixed quantum--classical studies of vibrational relaxation, where classical-bath treatments become especially unreliable for high-frequency modes; in modern terms, this breakdown is closely related to zero-point-energy leakage, where high-frequency vibrations spuriously transfer energy into softer modes~\citeS{egorovAdequacyMixedQuantum1999,habershonZeroPointEnergy2009}. 

The mixed quantum--classical scheme employed here faces a similar limitation: by treating slow phonons as classical fields without backaction (forces) from the quantum subsystem, it does not satisfy detailed balance either. At the same time, the present fast/slow partition mitigates the most severe leakage channel, because the high-frequency intramolecular phonons are not propagated as explicit classical trajectories but are traced out analytically through the variational polaron transformation. We therefore extract transport coefficients via the Green--Kubo formalism, where the current--current correlation function $C(\tau)$ naturally decays to zero and the mobility integral converges at short to intermediate times where the dynamics remain accurate, substantially reducing sensitivity to long-time artifacts. This, combined with variational polaron dressing that traces out fast phonons analytically, yields a tractable approach that captures the essential nonperturbative physics, as validated against exact benchmarks (Fig.~\ref{fig:dntt}a). Numerically exact methods such as tensor-network, path-integral, and hierarchical equations of motion approaches can in principle provide systematically improvable results on the same Hamiltonians, and represent a natural next step as algorithmic advances extend their reach to realistic system sizes.

A conceptually nearby framework is transient localization theory, where thermally activated low-frequency vibrations are represented as a quasi-static disorder landscape, often with Gaussian statistics, and the mobility is inferred from a transient localization length together with a relaxation-time approximation~\citeS{fratiniTransientLocalizationScenario2016}. 
The difference is that here the low-frequency phonons are sampled explicitly within the first-principles Hamiltonian and the transport coefficients are obtained directly from the Green--Kubo current correlation function, so no additional phenomenological relaxation time or localization-length ansatz is required. At the same time, the Hamiltonians developed here should provide a natural starting point for future transient-localization-style analyses.

\subsection*{Supplementary References}
\small
\begingroup
\renewcommand{\bibitem}[2][]{%
  \refstepcounter{SIref}%
  \par\noindent\hangindent=2.5em\hangafter=1%
  {[S\arabic{SIref}]}\label{SIref:#2}\enspace
}
\input{si_refs.tex}
\endgroup

\end{document}